\documentclass[aps,prb,10pt,superscriptaddress,amsmath,amssymb,twocolumn]{revtex4-2}
\usepackage{graphicx,bm,amsmath}
\usepackage[colorlinks]{hyperref}

\usepackage[bbgreekl]{mathbbol}
\usepackage{xcolor}
\usepackage[normalem]{ulem}
\usepackage{amssymb}
\usepackage{braket}
\usepackage{verbatim}
\usepackage{float}
\usepackage{todonotes}
\usepackage{appendix}

\def\doi{http://dx.doi.org/}

\newcommand{\be}{\begin{equation}}
\newcommand{\ee}{\end{equation}}
\newcommand{\bea}{\begin{eqnarray}}
\newcommand{\eea}{\end{eqnarray}}

\newcommand{\nep}{{\rm e}}

\newcommand{\tr}{\mathrm{Tr}}

\begin{document}
\title{Engineering of non-Hermitian interactions in digital qudit quantum simulators}

\author{Matteo M. Wauters}
\affiliation{Pitaevskii BEC Center and Department of Physics, University of Trento, Via Sommarive 14, I-38123 Trento, Italy}
\affiliation{INFN-TIFPA, Trento Institute for Fundamental Physics and Applications, Trento, Italy}
\author{Paolo Boschetto}
\affiliation{University of Trento, Via Sommarive 14, I-38123 Trento, Italy}
\author{Edoardo Ballini}
\affiliation{Pitaevskii BEC Center and Department of Physics, University of Trento, Via Sommarive 14, I-38123 Trento, Italy}
\affiliation{INFN-TIFPA, Trento Institute for Fundamental Physics and Applications, Trento, Italy}
\author{Alberto Biella}
\affiliation{Pitaevskii BEC Center and Department of Physics, University of Trento, Via Sommarive 14, I-38123 Trento, Italy}
\affiliation{INFN-TIFPA, Trento Institute for Fundamental Physics and Applications, Trento, Italy}
\author{Philipp Hauke}
\affiliation{Pitaevskii BEC Center and Department of Physics, University of Trento, Via Sommarive 14, I-38123 Trento, Italy}
\affiliation{INFN-TIFPA, Trento Institute for Fundamental Physics and Applications, Trento, Italy}

\begin{abstract}
    Non-Hermitian Hamiltonians are a fascinating class of many-body models that describe the effective dynamics of quantum systems interacting with the environment through particle, energy, or information exchange. Although their theoretical framework is well established, the controlled engineering of such Hamiltonians in the context of quantum simulations remains challenging, even more so when the non-Hermitian part describes a $k$-body interaction.
    Qudit quantum simulators offer a compelling framework to implement such models. 
    We theoretically investigate the dynamics of a one-dimensional chain of qudits undergoing hybrid unitary-projective evolution, where suitably designed measurements constrain the dynamics to a Zeno subspace.
    As we illustrate for the case of qutrits, within the Zeno subspace the dynamics is governed by an effective non-Hermitian Hamiltonian for an ensemble of pseudo-spins $1/2$, which can inherit non-Hermitian two-body interactions with the same connectivity as the full qutrit chain.  
    We derive an analytical relation linking the monitored qutrits' evolution to a desired target non-Hermitian Hamiltonian and validate the effective description through numerical simulations of a representative model. Our scheme provides a constructive route for the realization of a large class of interacting non-Hermitian many-body Hamiltonians in experimentally relevant multilevel quantum platforms, including trapped ions and superconducting circuits.
\end{abstract}

\maketitle

\section{Introduction}
Interest in non-Hermitian quantum physics has grown rapidly over the past decade, driven by the discovery of qualitatively new phenomena that have no Hermitian counterpart \cite{Ashida_2020,roccati2026}. These include exceptional points \cite{Meng2024,Minganti2019,Lakkaraju_2026}, non-Hermitian phase transitions \cite{Wang_pra2026,Rosso2023}, the non-Hermitian skin effect \cite{Zhang2022,Gohsrich2025}, and unconventional topological structures \cite{Bergholtz_RMP2021,Fleckenstein2022,Jia2025}. 
Many of these effects have been explored experimentally across a variety of platforms, including photonic systems \cite{Xiao2025}, electrical circuits \cite{Guo2024,Chen_2026}, cold atoms \cite{Wang_arx2026,Luo2024}, and superconducting quantum devices \cite{Song2024,Wang2024}, where gain and loss processes or measurement backaction can be precisely controlled. In several of these architectures, effective non-Hermitian Hamiltonians emerge as the generators of conditional dynamics, for example, in postselected quantum trajectories associated with monitored open systems \cite{Turkeshi_PRB2021,minganti2024openquantumsystems,fazio2025manybodyopenquantumsystems}.

Despite this progress, the controlled engineering of general non-Hermitian Hamiltonians remains a nontrivial task. While simple non-Hermitian contributions—such as single-particle gain and loss terms—arise naturally and can be implemented with a high degree of control in many experimental platforms~\cite{Lee_PRX2014,Labouvie_PRL2016,Morsch2018,Sponselee_2019,Cao_PRL2020}, going beyond this regime to engineer more general non-Hermitian Hamiltonians is significantly more challenging.
 In particular, constructing arbitrary non-Hermitian operators tailored to specific target models typically requires the implementation of fine-tuned interactions between the system and multiple ancillary qubits, increasing the circuit complexity~\cite{Okuma_PRB2022,Lin_npjQI2022,Kamakari_PRXQ2022, Geier_PRXQ2022, Liu_PRA2023,Shen_NatComm2025,Agarwal_PRA2026}. 
Yet, such interacting non-Hermitian Hamiltonians are expected to host rich and largely unexplored many-body phenomena~\cite{Yamamoto2019,Fossati2023,Turkeshi2022,Biella2021manybodyquantumzeno}, making them a compelling target for quantum simulation protocols~\cite{roccati2026,shen_arxiv2026}.

In this work, we propose an alternative hardware-efficient framework for engineering interacting non-Hermitian Hamiltonians in digital quantum simulations, leveraging measurement-induced quantum Zeno (QZ) dynamics~\cite{misra1977thezeno, peres80zeno, Itano_PRA1990, FACCHI_PLA2000, facchi2001from, facchi2002quantum, signoles2014confined, snizhko2020quantumzeno}, where the ancilla qubits are substituted with auxiliary levels within the local Hilbert space of {\em qudits}~\cite{Wang_2020}. As a minimal setup, we consider a one-dimensional chain of qutrits undergoing hybrid evolution where the unitary dynamics is punctuated with projective measurements. By appropriately designing the measurement protocol, the system dynamics is constrained to a Zeno subspace in which the effective evolution of the first two local levels (an effective spin-$1/2$ system) is governed by a non-Hermitian Hamiltonian~\cite{Barge_2026}. Importantly, the structure of this effective Hamiltonian can include anti-Hermitian interaction terms inherited from the underlying unitary dynamics of the full qutrit chain.

Our approach leverages two types of recent experimental progress: the control of multilevel quantum systems and mid-circuit measurements. On the one hand, qudits~\cite{Wang_2020} (and in particular qutrit networks) have become increasingly accessible across several platforms, where coherent manipulation, high-fidelity measurements, and tunable interactions can be achieved with a high degree of precision~\cite{Low_PRR2020, Chi_NatComm2022, Ringbauer_2022, Liu_PRX2023, Fischer_PRXQ2023, Burshtein_PRR2026}. A prominent example is provided by trapped ions, where each atom intrinsically hosts multiple energy levels that can be selectively addressed and controlled~\cite{Ringbauer_2022,Hrmo_NatComm2023}. These systems offer a flexible and scalable architecture in which both unitary dynamics and measurement protocols can be engineered with fine temporal and spatial resolution, making them especially well suited for the implementation of the scheme proposed here.
On the other hand, mid-circuit measurements are moving into the spotlight as an essential tool in digital quantum computing, with applications ranging from quantum error correction~\cite{Shor_PRA1995,Steane_PRL1996}, mitigation~\cite{Cai_RMP2023,Ballini_quantum2025} and measurement-based quantum computing~\cite{Raussendorf2001} to measurement-induced phase transitions~\cite{Li2018,Skinner_PRX2019,Turkeshi_PRB2021}. They have been successfully implemented in several qubit platforms~\cite{Noel_NatPhys2022,Rudinger_PRApp2022,Deist_PRL2022,Koh_NatPhys2023,Graham_PRX2023,Yu_PRR2025}.
Combining mid-circuit measurements with qudit platforms is more challenging than their qubit counterpart, e.g., because one needs to maintain phase coherence of the idle levels during single-level readout~\cite{Ringbauer_2022}. Nonetheless, first experiments have already demonstrated encouraging results~\cite{Chi_APLPhot2026}.

The present framework provides an explicit constructive route to realize a broad class of non-Hermitian models. Through the well-established QZ formalism~\cite{Facchi_2008,Geier_PRXQ2022}, we show how to engineer the microscopic unitary dynamics of the qudit chain to generate a desired effective non-Hermitian Hamiltonian. 
In addition, we support this analytical construction with numerical simulations of the monitored dynamics.
We demonstrate the validity of the effective description in a representative model with anti-Hermitian density–density coupling. We show that the unitary-projective qutrit dynamics correctly captures its rich physics, including a phenomenology hinting at boundary time crystals~\cite{Iemini_PRL2018} and synchronization~\cite{Zhirov2006,Schmolke_PRL2022} effects.

Our results highlight the flexibility of the method and its potential for exploring interacting non-Hermitian many-body physics.
More broadly, they establish a complementary method to existing tools for the controlled engineering of non-Hermitian quantum dynamics with minimal resource overhead, in terms of the required number of ancillary degrees of freedom and unitary operations, opening the door to the systematic investigation of interacting non-Hermitian phases in experimentally relevant quantum platforms.

The rest of this article is organized as follows. In Sec.~\ref{sec:framework}, we set up the theoretical framework of our proposal and derive the correspondence between the post-selected (PS) dynamics and its effective non-Hermitian description. In Section~\ref{sec:results}, we provide numerical benchmarks on a selected test case, a spin-$1/2$ system with 2-body dissipation, while we summarize our findings and present future outlooks in Sec.~\ref{sec:conclusion}.
The Appendix~\ref{app:oddN} highlights the dependence of the steady state degeneracy of the benchmark model on its system-size parity. Finally, we hint at the construction of different classes of non-Hermitian models in Appendix~\ref{app:extra_models}, in particular a non-Hermitian quantum East model and one implementing the non-Hermitian skin effect.

\section{Framework}\label{sec:framework}
Let us start by reviewing the formalism to derive an effective non-Hermitian Hamiltonian within the QZ framework, specialized here to the qudit platforms we consider.
Similar derivations can be found, e.g., in Refs~\cite{Facchi_2008,Geier_PRXQ2022}.

In the context of performing weak measurements to induce an effective non-Hermitian dynamics through the QZ effect, a standard setup consists of coupling the qubit(s) one wishes to weakly monitor to an ancillary qubit and performing a projective readout on the latter~\cite{Stannigel_PRL2014,Noel_NatPhys2022,Wauters_PRB2025,Guttel2026}.
Leveraging the larger local Hilbert space within which the qubits are usually embedded offers an alternative paradigm for implementing quantum-simulation protocols targeting non-Hermitian dynamics. 
Instead of using ancillary degrees of freedom, we consider {\em qudit} quantum information carriers and split their local Hilbert spaces of dimension $d>2$ into a computational subspace and an auxiliary level. 
A negative readout of the auxiliary level projects the system's state onto the computational subspace. Hence, preparing the system in the computational subspace and frequently monitoring the auxiliary levels drives the system into the QZ regime, inducing an effective non-Hermitian dynamics on the computational subspace~\cite{Barge_2026}.
Compared to the qubit approach, this method has two main advantages: first, it can be immediately applied to any qudit size $d$, to induce non-Hermitian dynamics on an effective spin-$\frac{d-1}{2}$ system.
Second, it reduces the number of entangling gates needed since no further coupling with ancillary degrees of freedom is involved, as we will detail below.

To be concrete, let us consider a chain of $N$ qutrits ($d=3$), where the computational levels $\{\ket{0}_n, \ket{1}_n\}$ (with $n=1,\dots,N$) define an ensemble of qubits, and we alternate unitary evolution $U(\Delta t)$ with projective measurements on the auxiliary levels $\{\ket{2}_n\}$. 
In the following, we denote with $P$ and $Q$ the projectors on the computational and auxiliary subspaces, respectively.
For a clearer derivation, we assume that the unitary operator $U(\Delta t) $ is generated by a tunable 2-local Hamiltonian acting on the full qutrit Hilbert space. For the implementation on a specific hardware, such as trapped-ion \cite{Ringbauer_2022} or superconducting qudits \cite{Liu_PRX2023}, this evolution can be further decomposed into elementary gates.
Moreover, we assume initialization of the system in a state $\ket{\psi_0}=P\ket{\psi_0}$ fully within the computational space.

The above evolution of the density matrix $\rho(t)$ is then described as a discrete-time channel, where a projective measurement follows the unitary evolution, selecting the subspace $P$ or $Q$, 
\begin{equation}\label{eq;proj_evo_full}
    \rho(t+\Delta t) = P \nep^{-iH\Delta t}\rho(t) \nep^{iH\Delta t}P + Q\nep^{-iH\Delta t}\rho(t) \nep^{iH\Delta t}Q \ .
\end{equation}
In the QZ regime, we can neglect the population in the auxiliary $Q$-subspace at time $t$, meaning that $P\rho(t)P =\rho(t)$, as we assume that the density matrix has been projected onto the $P$-subspace by the previous round of measurements.
In practice, this is reached when the measurement interval $\Delta t$ is small. 
Introducing the notation $P X P\equiv X_{PP}$, $P X Q\equiv X_{PQ}$, etc., a second-order expansion of Eq.~\eqref{eq;proj_evo_full} in $\Delta t$ leads to
\begin{align}
    \rho_{PP}(t+\Delta t) = & \rho_{PP}(t) - i\Delta t\left[H_{PP}, \rho_{PP}(t) \right] \nonumber \\
    + &\frac{\Delta t^2}{2}\left( 2H_{PP} \rho(t)H_{PP}-\left\{ H_{PP}, \rho_{PP}(t) \right\} \right) \nonumber \\
    -& \frac{\Delta t^2}{2} \left\{ H_{PQ}H_{QP},\rho_{PP}(t) \right\} + \mathcal{O}(\Delta t^3) \ .\label{eq:ev_rhoPP_2ndOrder}
\end{align}
Focusing on $\rho_{PP}(t+\Delta t)$ only is equivalent to post-selecting the measurement outcomes where no qudit has been found in the auxiliary state.

The first two lines of Eq.~\eqref{eq:ev_rhoPP_2ndOrder} correspond to the expansion of the von Neumann equation associated with the Hamiltonian $H_{PP}$ alone, thus describing a unitary evolution up to order $\Delta t^2$. 
The last line is governed by terms of $H_{PQ}$ that are independent of $H_{PP}$. We can interpret it as being derived from an effective non-Hermitian Hamiltonian, which can formally be added to the first-order expansion in the first line. Factoring out one order of $\Delta t$, this non-Hermitian contribution is given by 
\begin{equation}\label{eq:nh_approx0}
    H_{\rm n.h.} \simeq -i \frac{\Delta t}{2} H_{PQ}H_{QP} \ .
\end{equation}

For a better understanding, it is useful to explicitly expand Eq.~\eqref{eq:nh_approx0} on the computational basis $\{\ket{\bf s}\} \cup \{ \ket{\bf q}\}$, where the two sets correspond to the states stabilized by either $P$ or $Q$, 
\begin{equation}\label{eq:nh_approx}
    H_{\rm n.h.} \simeq -i \frac{\Delta t}{2}\sum_{{\bf s},{\bf s'}} \sum_{\bf q} \bra{\bf s} H_{PQ} \ket{\bf q} \bra{\bf q} H_{QP} \ket{\bf s'}\ .
\end{equation}
The non-Hermitian processes we can describe are therefore those connecting two states in the computational subspace $P$ through the virtual occupation of an auxiliary state $\ket{\bf q}$ in subspace $Q$.

Equations~\eqref{eq:nh_approx0} and~\eqref{eq:nh_approx} have been obtained through a post-selection on the projective-readout outcomes. It is important to estimate its failure rate, i.e., the probability of finding at least one qudit in the auxiliary level.
Expanding again Eq.~\eqref{eq;proj_evo_full} to second order, with the assumption $\rho(t)=\rho_{PP}(t)$, we get that the probability of projecting the density matrix onto the $Q$-subspace is 
\begin{equation}
    p_{\rm err} (\Delta t)= \tr \rho_{QQ}(t+\Delta t) = \Delta t^2 \tr\left( \rho_{PP}(t) H_{PQ}H_{QP}\right) \ .
\end{equation}
This sets the main limitation of our protocol: frequent measurements ($\Delta t \ll 1$) prevent the system from leaking out into the auxiliary subspace thanks to the QZ effect, but, at the same time, suppress the strength of the effective non-Hermitian Hamiltonian given in Eq.~\eqref{eq:nh_approx0}. 
A realistic implementation of this protocol, therefore, is reduced to situations where the non-Hermitian terms are relatively small perturbations on top of the Hermitian ones, so that we can properly understand the overall dynamics within the framework of the QZ effect.
However, we will show below that the agreement between the PS dynamics and the effective non-Hermitian one in Eq.~\eqref{eq:nh_approx} can be surprisingly good even at strong $P$-$Q$ coupling.

\subsection{Building blocks of effective non-Hermitian Hamiltonian}\label{ssec:blocks}

So far, the derivation has been rather general, and similar results can be found, for instance, in Refs.~\cite{Geier_PRXQ2022,Stannigel_PRL2014}.
We now specialize to the qutrit chain we consider in the rest of this article to understand which terms can be derived from simple one- and two-body terms in the full Hamiltonians. 

\begin{figure}
    \centering
    \includegraphics[width=\columnwidth]{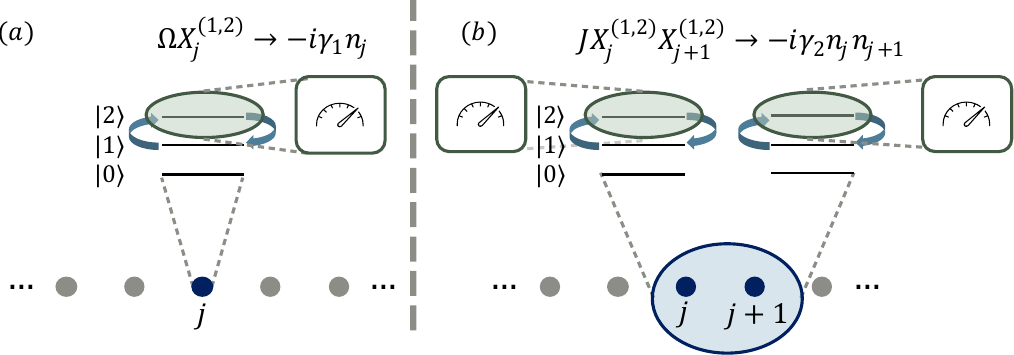}
    \caption{(a) Local dissipative term and (b) non-Hermitian interaction. Every dot represents a qutrit. The two terms are realized by coupling the $P$ and $Q$ subspaces through $X_j^{(1,2)}$ and $X_j^{(1,2)}X_{j+1}^{(1,2)}$, respectively. The effective non-Hermitian dynamics is achieved by independent projective readout of the state $\ket{2}$ of each qutrit, following the unitary evolution.}
    \label{fig:non-hermitian_terms}
\end{figure}

In general, we will consider Hamiltonians with nearest-neighbour interactions of the form
\begin{align}
    \label{eq:Hfull}
    H_{\rm full} &=  \sum_j \sum_{\alpha, \beta} \Omega_{\alpha,\beta} P_j^{(\alpha,\beta)} \nonumber \\
    &+ \sum_j\sum_{\alpha,\beta} \sum_{\alpha',\beta'} J^{\alpha,\beta}_{\alpha',\beta'} P_j^{(\alpha,\beta)}  P_{j+1}'^{(\alpha',\beta')}  \ ,
\end{align}
where $j$ labels the qutrits, $P=X, Y,Z$ are spin-$\frac{1}{2}$ Pauli operators embedded in the larger qutrit space and acting on the pairs of levels $\alpha,\beta=0,1,2$. $\Omega$ and $J$  are the strengths for single-qutrit rotations and two-qutrit interaction terms, respectively. 
Such operations are easily derived from the natural gates realized, e.g., in universal trapped-ion qudit hardware ~\cite{Ringbauer_2022}. 
Further terms can be incorporated in a straightforward manner, but we find the present structure to be already sufficiently rich for our purposes.  

For simplicity, in the following we moreover assume all terms in Eq.~\eqref{eq:Hfull} to be site-independent. 
In addition, for the time being, we ignore terms where the Pauli strings act only within the $P$-subspace ($\alpha,\beta \in \{0,1\}$), as such terms are directly translated into the corresponding qubit operators. 
Rather, in this section, we provide three examples of effective non-Hermitian terms that can be obtained with the above protocol.
Figure~\ref {fig:non-hermitian_terms} sketches the first two of such examples, which enter the benchmark model studied in Sec.~\ref{sec:results}.

\paragraph{Local dissipation:}
In the simplest scenario, the coupling between the $P$ and $Q$ subspaces happens through local driving terms such as 
\begin{equation}
    \label{eq:HPQlocal}
    H_{PQ} = \Omega \sum_j X^{(1,2)}_j \ .
\end{equation}
This Hamiltonian induces the transition $\ket{1} \leftrightarrow \ket{2}$ on individual qutrits, while it acts trivially on the state $\ket{0}$. Thus, Eq.~\eqref{eq:nh_approx} becomes 
\begin{eqnarray}
    H_{\rm n.h.} \simeq -i \frac{\Omega^2 \Delta t}{2} \sum_j \ket{1}\bra{1}_j = -i \frac{\Omega^2\Delta t}{2} \sum_j n_j \ , 
\end{eqnarray}
where $n_j=(1-\sigma^z_j)/2$ is the local projection on the $\ket{1}$ state. 
Such a non-Hermitian term is relatively standard. E.g., this same setting has been recently considered in Ref.~\cite{Barge_2026} for a single superconducting qutrit, and its implementation in qubit platforms is also straightforward~\cite{Geier_PRXQ2022,Noel_NatPhys2022}. 

\paragraph{Non-Hermitian density-density interaction:}
Of greater interest in our context is the possibility of generating genuine non-Hermitian \emph{interactions} by combining measurements with the native couplings of the underlying qutrit system.
For instance, we may consider terms akin to generalized M{\o}lmer-S{\o}rensen gates for trapped ions~\cite{Sorensen_PRA2000, Ringbauer_2022}, 
\begin{equation}
    H_{PQ} = J \sum_j X^{(1,2)}_j X^{(1,2)}_{j+1} \ .
\end{equation}
Now, the matrix elements $\bra{\bf s} H_{QP}\ket{\bf q}$ in Eq.~\eqref{eq:nh_approx} are nonzero only when the two states $\ket{\bf s}$ and $\ket{\bf q}$ differ by a correlated spin flip $\ket{11} \leftrightarrow \ket{22}$ on neighbouring qutrits. Hence, the terms appearing in Eq.~\eqref{eq:nh_approx} take the form
\begin{eqnarray}
    H_{\rm n.h.} \simeq -i\frac{J^2 \Delta t}{2}\sum_j n_j n_{j+1} \ .
\end{eqnarray}
Crucially, embedding the qubit ensemble in a qutrit system allows for obtaining effective non-Hermitian interactions without extra overhead other than the mid-circuit readout of the auxiliary levels. In contrast, in a qubit setup, one would need to implement at least a three-body interaction involving two computational and one ancillary qubit.

\paragraph{Combination of multiple terms:}
Up to now, we have considered Hamiltonians $H_{PQ}$ consisting of a single type of $P$-$Q$ coupling. This does not have to be the case, and we can combine the two terms we have seen before
\begin{eqnarray}\label{eq:H_PQ_com}
    H_{PQ} = \Omega \sum_j X^{(1,2)}_j + J \sum_j X^{(1,2)}_j X^{(0,1)}_{j+1} \ .
\end{eqnarray} 
Notice also that in this example the $XX$-interaction involves different pairs of levels of the two qutrits, a situation that can easily be achieved, e.g., in state-of-the-art trapped-ion hardware~\cite{Ringbauer_2022}.

The second-order expansion in Eq.~\eqref{eq:nh_approx} now contains three different virtual paths, corresponding to the different combinations of the two terms in Eq.~\eqref{eq:H_PQ_com}
\begin{eqnarray}
    & \ket{1,s} \xrightarrow{X_j^{(1,2)}X_{j+1}^{(0,1)}} \ket{2,\bar{s}} \xrightarrow{X_j^{(1,2)}X_{j+1}^{(0,1)}} \ket{1,s} \ , \nonumber \\
    & \ket{1,s} \xrightarrow{X_j^{(1,2)}} \ket{2,s} \xrightarrow{X_j^{(1,2)}} \ket{1,s} \ , \nonumber \\
    & \ket{1,s} \xrightarrow{X_j^{(1,2)}X_{j+1}^{(0,1)}} \ket{2,\bar{s}} \xrightarrow{X_j^{(1,2)}} \ket{1,\bar{s}} \ , \nonumber \\
\end{eqnarray}
(plus the path that inverts the order of application of $X_j^{(1,2)}X_{j+1}^{(0,1)}$ and $X_j^{(1,2)}$). 
These lead to two qualitatively different contributions in the effective Hamiltonian,
\begin{equation}
    H_{\rm n.h.}\simeq -i\frac{(\Omega^2 +J^2)\Delta t}{2}\sum_j n_j + -iJ \Omega \Delta t\sum_j n_j \sigma^x_{j+1}  \ .
\end{equation}
The first term, generated by the first two of the above paths, describes the rather straightforward local dissipation as also obtained from Eq.~\eqref{eq:HPQlocal}. 
On top of that, $H_{\rm n.h.}$ contains a non-Hermitian spin flip facilitated by the occupation of the neighbouring site. 
As we discuss in App.~\ref{app:extra_models}, this type of term permits the realization of non-Hermitian variants of quantum East models \cite{Pancotti_PRX2020,das_2026}. 

As these examples illustrate, the toolbox obtained already with only single- and two-qutrit couplings enhanced by mid-circuit measurements is quite rich, permitting a large freedom in the design of effective non-Hermitian dynamics.
\begin{figure*}
    \centering
    \includegraphics[width=\linewidth]{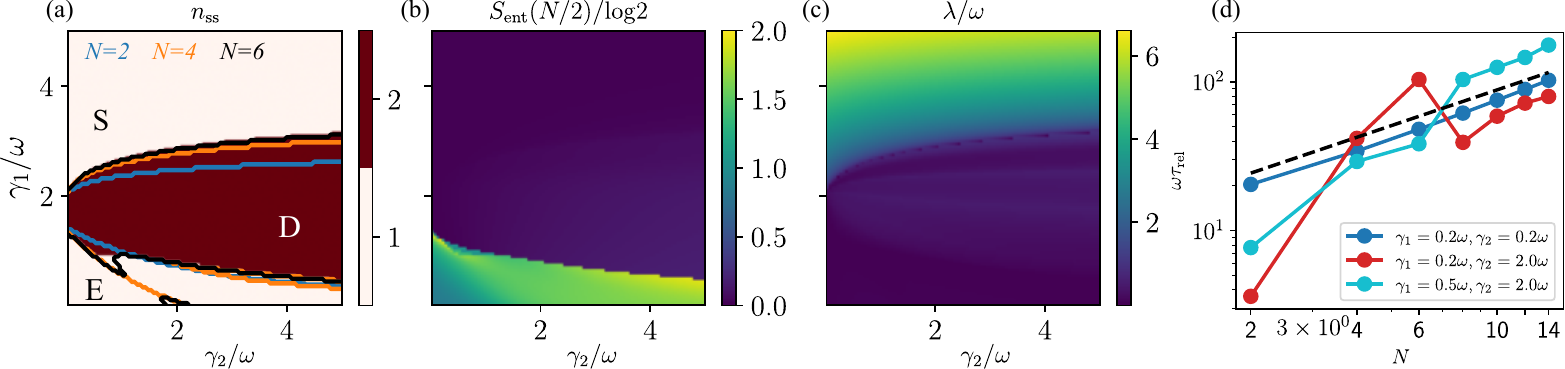}
    \caption{Ground state degeneracy (a), entanglement entropy (b), and relaxation rate (c) for the effective non-Hermitian qubit model with $N=8$. In panel (a), the solid lines indicate the boundaries of the D phase for different system sizes. These quantities let us identify $3$ different phases of the system, defined by the strengths of $\gamma_1$ and $\gamma_2$: degenerate phase (D; double degeneracy of the steady state, low entropy, and low relaxation rate), entangled phase (E; single steady state, high entropy, and low relaxation rate), and standard phase (S; single steady state, low entropy, and high relaxation rate).
    (d) Inverse of the relaxation rate in the E phase as a function of system size $N$. Despite the small values of $N$, a clear trend is visible, compatible with a $N^{0.8}$ scaling (dashed black line).}
    \label{fig:phase_diagram_N8}
\end{figure*}

\section{Numerical benchmark: non-Hermitian density--density coupling}\label{sec:results}

We now focus our discussion on a specific benchmark model to showcase the potential and limitations of the present framework in a concrete setting. 
We focus on a spin-$\frac{1}{2}$ system, where the Hermitian part consists only of Rabi oscillations with a uniform frequency $\omega$, while the non-Hermitian contribution describes single-body losses and density-density interaction. 
Importantly, the genuine non-Hermitian interactions allow for the emergence of many-body effects beyond the single-particle non-Hermitian physics more commonly studied~\cite{zhang2022reviewnonhermitianskineffect,Wang_pra2026,Bergholtz_RMP2021,Luo2024,Lee_PRX2014}.

The considered model Hamiltonian reads 
\begin{equation}\label{eq:H_zeni}
    H_{\rm eff} = \omega \sum_j \sigma^x_j -i\gamma_1 \sum_j n_j -i\gamma_2 \sum_j n_j n_{j+1} \ .
\end{equation}
This model has been theoretically studied in Ref.~\cite{zeni2025theorycorrelatedquantumzeno} for $N=2$ with a Gutzwiller mean-field approximation, showing a non-trivial phase diagram due to the interplay between the unitary dynamics and the dissipative channels.
Besides its interesting physical phenomenology, the presence of genuine non-Hermitian interaction makes this model challenging to implement in a quantum-simulation protocol, and thus an ideal target for our proposal.

Using the building blocks discussed in Sec.~\ref{ssec:blocks}, we can identify a microscopic qutrit Hamiltonian that generates this effective non-Hermitian model, 
\begin{equation}
    \label{eq:Hfull_zeni}
    H_{\rm full} = \omega\sum_j X^{(0,1)}_j + J \sum_j X^{(1,2)}_j X^{(1,2)}_{j+1} + \Omega \sum_j X^{(1,2)}_j \ ,
\end{equation}
Using Eq.~\eqref{eq:nh_approx}, we find the parameters of the effective and microscopic model are related as $\gamma_1 = \Omega^2 \Delta t/2 $ and $\gamma_2 = J^2 \Delta t/2 $.
Contrary to the example (c) in the previous section, mixed terms $\sim J \Omega$ leave one of the qutrits in the state $\ket{2}$ and are thus projected out. 
Terms that act only on the computational subspace $P$, instead, are not modified by the measurements and post-selection, so we can immediately substitute $X^{(0,1)}_j \to \sigma^x_j$.

In the following, we first describe the steady-state phenomenology of the model set by Eq.~\eqref{eq:H_zeni} and identify interesting physical regimes, before investigating the regimes of validity of this effective model. 

\subsection{Phase diagram of the effective non-Hermitian model}
In a non-Hermitian Hamiltonian, eigenvalues $\{\varepsilon_i\}$ are typically complex, their (negative) imaginary part being associated with dissipation processes. The steady-state is then the Hamiltonian's {\em right eigenvector} with the imaginary part closest to zero, meaning that it has the slowest decay rate. When this condition applies to multiple eigenvalues, i.e., $\exists\ \varepsilon_i,\varepsilon_j\ | \ \Im \varepsilon_i = \Im \varepsilon_j = \min_{\{\varepsilon_n\}} \Im \varepsilon_n$, there are multiple steady states, possibly preventing the system from relaxing to equilibrium. Interestingly, the model under study displays a different phenomenology in this regard depending on whether $N$ is even or odd. 

To characterize the phase diagram, we monitor three quantities: the steady-state degeneracy $n_{\rm ss}$, its half-chain entanglement entropy $S_{\rm ent}(N/2)$, and the relaxation rate $\lambda$.
If the steady state is degenerate, i.e., two or more eigenvalues have the same (smallest) imaginary part, we define $S_{\rm ent}(N/2)$ as the average of the entanglement entropy of each stationary state. 
The relaxation rate, instead, is $\lambda = \min_{n\notin {\rm ss}} |\Im \varepsilon_n-\Im \varepsilon_{\rm ss}|$, where $\varepsilon_n$ does not belong to the steady-state manifold ${\rm ss}$.

Figure~\ref{fig:phase_diagram_N8}(a-c) displays the phase diagram of Eq.~\eqref{eq:H_zeni} for $N$ even (see Appendix~\ref{app:oddN} for $N$ odd); we can identify three distinct regions, which we denote the entangled (E), degenerate (D), and standard (S) phases.
The E region is dominated by the interplay between the unitary evolution and non-Hermitian interaction strength $\gamma_2$. It is characterized by a unique entangled steady state and a relaxation rate that decreases with a power law in $N$. 
The boundary of the E region changes with $N < 8$ (see contour lines for different system sizes in Fig.~\ref{fig:phase_diagram_N8}(a)) but then converges after $N\ge 8$. 
The entanglement entropy of the steady state in the E phase is markedly larger than that in the rest of the phase diagram, see Fig.~\ref{fig:phase_diagram_N8}(b). 
Furthermore, it displays a weak sub-linear dependence on the system size (not shown). This behavior is reminiscent of the $\log N$ scalings that are known to appear in measurement-driven entanglement phase transitions~\cite{Skinner_PRX2019, Turkeshi_PRB2021, Koh_NatPhys2023, Noel_NatPhys2022}, but full characterization of the entanglement dynamics in the effective non-Hermitian model requires larger system sizes and is left for future work.
The increasingly slow equilibration is clearly shown in panel (d), where small-size data suggest that the relaxation time $\tau_{\rm rel}=1/\lambda$, the inverse of the relaxation rate, increases as a power law of $N$, $\omega \tau_{\rm rel} \propto N^\beta$, with $\beta \simeq 0.8$ only weakly varying in the E region.
The apparent divergence of the relaxation rate with system size suggests that the E phase might support boundary time crystals~\cite{Iemini_PRL2018, Piccitto_PRB2021}, which will be an interesting research question for future studies.

When all three terms in the Hamiltonian are relevant, i.e., when the one-body dissipation term $\gamma_1$ is comparable with $\gamma_2$ and $\omega$, the system enters the D phase, exhibiting two weakly entangled steady states. After a transient dynamics, on a timescale set by the relaxation rate, the system remains out of equilibrium, constantly oscillating between them.
Here, the size-dependence of the entanglement entropy emerges from the dynamics: while both steady states have low entanglement, their superposition induced by the non-Hermitian evolution oscillates between large and low entanglement, see the following paragraphs for an example. The time-averaged entanglement entropy of the resulting non-equilibrium steady state matches the weak $N$-dependence of the E phase.

Finally, the S phase arises when the dynamics is dominated by the one-body dissipation rate $\gamma_1$, and the system quickly relaxes towards a unique steady state that becomes fully separable in the $\gamma_1 \to \infty$ limit.

\subsection{Phenomenology and dynamics of the effective model}

\begin{figure*}
    \centering
    \includegraphics[width=\linewidth]{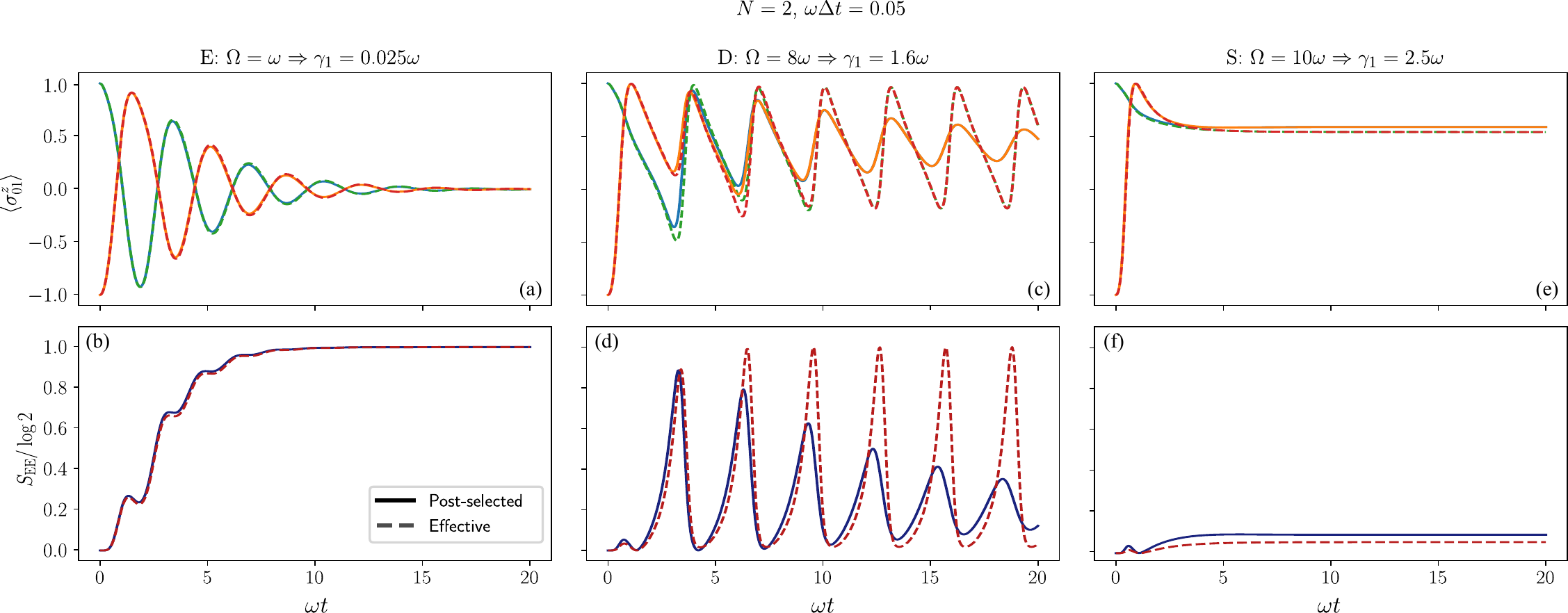}
    \caption{On-site magnetization dynamics (top panels) and half-chain entanglement entropy (bottom panels) in the three phases of the system for $N=2$: Entangled (a-b), Degenerate (c-d), and Standard (e-f). Solid lines mark the PS dynamics on the qutrit system, while dashed lines mark the target effective model. Different colors (blue/orange and green/red) in the top panels indicate the left/right spin in the dimer. In all panels, $\Delta t \omega =0.05$ and $J=8\omega $, equivalent to  $\gamma_2 = 1.6 \omega$, while $\gamma_1$ increases from left to right.
    The PS dynamics correctly capture the target non-Hermitian one, with the exception of the D phase, where they match in the transient regime while the steady state of the PS simulation resembles more that of the S phase.}
    \label{fig:N2_phenomenology}
\end{figure*}

Digital quantum simulators are typically more naturally employed for the study of out-of-equilibrium dynamics than of steady states.  
To underscore the potential of the protocol described in Sec.~\ref{sec:framework}, we characterize the non-unitary dynamics in the three regimes through the evolution of a local observable, the ${\bf z}-$magnetization, as well as the half-chain entanglement entropy. 
We compare the time evolution generated by the target model given in Eq.~\eqref{eq:H_zeni} with the PS dynamics on the full qutrit system, i.e., we simulate Eq.~\eqref{eq;proj_evo_full} for interleaving the unitary time evolution generated by the Hamiltonian in Eq.~\eqref{eq:Hfull_zeni} with projective measurements of the ancilla level. A more detailed analysis is relegated to Sec.~\ref{ssec:comparison}. Remarkably, the interesting phenomenology characterizing the phase diagram sharply appears already for $N=2$, as we report in Fig.~\ref{fig:N2_phenomenology} for a system initialized as $\ket{\psi_0}=\ket{01}$.

In the E phase, see Fig.~\ref{fig:N2_phenomenology}(a-b), the system relaxes towards the unique steady state after a relatively long transient; during this relaxation, the exact dynamics depend on the initial state. For the chosen $\ket{\psi_0}$, the two spins display damped oscillations that maintain the initial $\pi$ phase shift.
At the same time, correlations build up and reach the stationary value corresponding (for $N=2$) to a maximally entangled state. 
Here, the PS qutrit dynamics follow almost exactly the target non-unitary evolution, despite the coupling $J$ being the dominant term in the qutrit Hamiltonian and $\Delta t J=0.4$ not being excessively small. 

In the D phase, see panels (c-d), the transient toward a nonequilibrium steady state is somewhat shorter, and the two spins lock their phases regardless of the initial state. Persistent oscillations in both the magnetization and the entanglement entropy survive thanks to the steady-state degeneracy. 
The PS qutrit simulations are faithful at short to intermediate times, permitting to reproduce the target features up to time scales where the effective model has already relaxed to its non-equilibrium stationary state. 
At longer times, the PS simulations show a relaxation towards a low-entangled stationary regime. While the oscillations in both magnetization and entropy at these times are damped, they do capture the constant offset and the periodicity seen in the target model.

Finally, the S phase, panels (e-f), corresponds to the intuitive picture of a strongly dissipative dynamics, where the system quickly relaxes towards a trivial steady state. Although the effective couplings in the Hamiltonian Eq.~\eqref{eq:H_zeni} are significantly beyond the perturbative regime with respect to the Hermitian energy scale $\omega$, and despite $\Delta t \Omega =0.5$ being rather large, the PS dynamics matches again well the target model, with only a minor systematic error.  

\begin{figure}
    \centering
    \includegraphics[width=\linewidth]{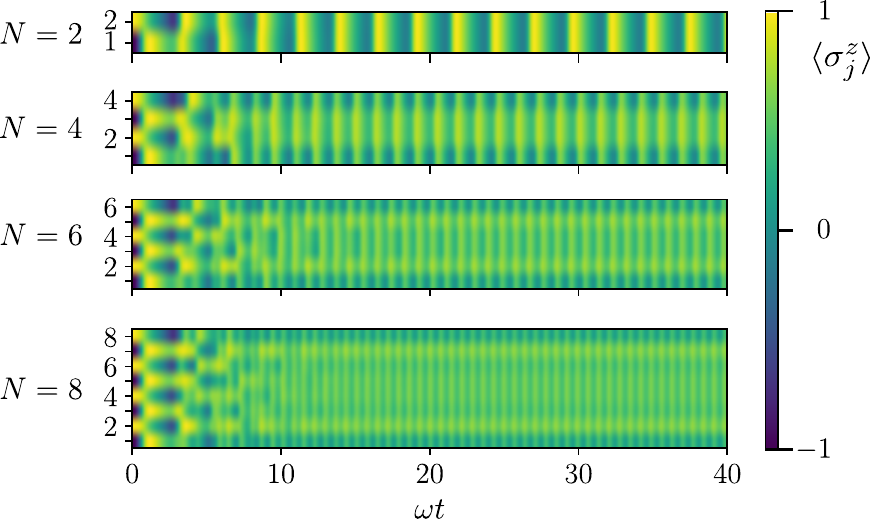}
    \caption{Synchronization of the qubit oscillations in the D phase, for different system sizes $N$. Data correspond to the effective spin-$\frac{1}{2}$ dynamics with couplings $\gamma_1=1.25\omega$ and $\gamma_2=2.45\omega$, with $\omega \Delta t=0.1$. 
    The synchronous oscillations reached after a rather short transient show a frequency that increases with system size and a position-dependent amplitude.
    In contrast, the time scale for the onset of synchronization is relatively independent of system size.  
    }
    \label{fig:D_vsN}
\end{figure}

When increasing the system size, the qualitative features of the three phases persist, and the D phase remains the hardest to capture with the PS qutrit dynamics. However, two noticeable modifications occur: {\em (i)} the relaxation time in the E regime increases with $N$, see Fig.~\ref{fig:phase_diagram_N8}(d); {\em (ii)} the amplitude of the oscillations in the D phase becomes spatially modulated, and their frequency increases with $N$, as exemplified by Fig.~\ref{fig:D_vsN}.
This synchronization effect~\cite{Zhirov2006,Schmolke_PRL2022} is quite remarkable and occurs on a time scale independent of both the initial state and the system size.
The size independence of the synchronization time can be appreciated by looking at the initial oscillations of the local magnetization in Fig.~\ref{fig:D_vsN}, which---up to the spatial increase---are identical in all four panels.

\subsection{Comparison with the post-selected dynamics}\label{ssec:comparison}
We now study in more detail the faithfulness and limitations of the PS qutrit dynamics in simulating the target non-Hermitian model. Here, we have to distinguish two sources of errors: the first is the intrinsic accuracy of the protocol, i.e., how well the PS Eq.~\eqref{eq;proj_evo_full} approximates the effective dynamics generated by Eq.~\eqref{eq:nh_approx}. The second is the bane of any PS or no-click dynamics: the exponential overhead in time and system size needed to extract the desired event out of all possible measurement outcomes.

\begin{figure*}
    \includegraphics[width=\linewidth]{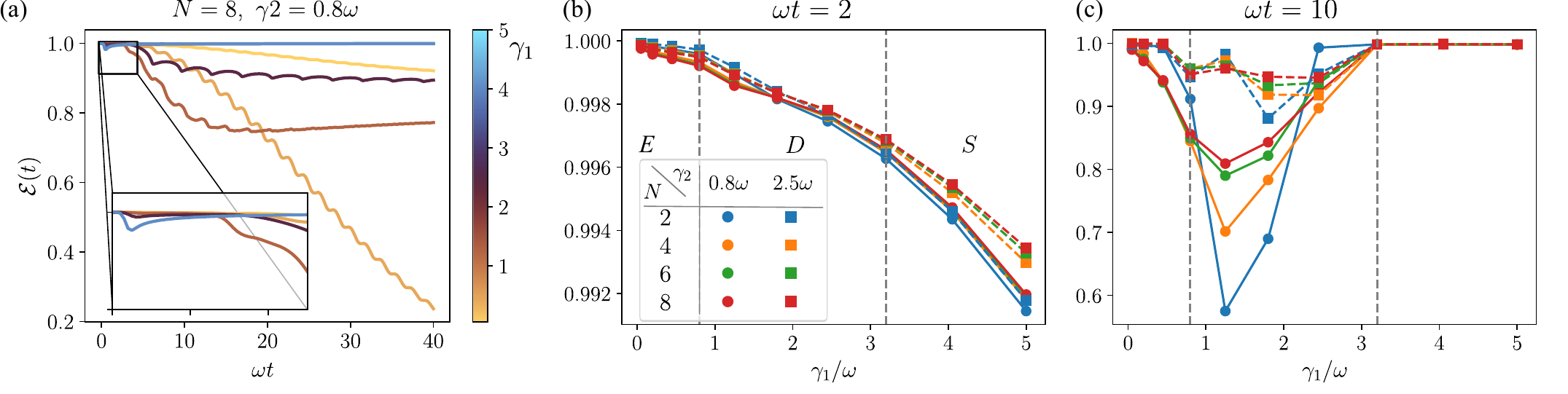}
    \caption{(a) Approximation ratio vs.\ time with varying one-body dissipation rate $\gamma_1$ (color-coded). We do not find a single universal behaviour, as the accuracy of the optimization at long time strongly depends on the specific parameter set. We see an excellent agreement at short times instead (inset) for the entire parameter range.
    (b) Approximation ratio $\mathcal{E}(t)$ vs.\ $\gamma_1$ at a fixed short time ($\omega t=2$), for different system sizes and two-body dissipation rate $\gamma_2$. The quality of the approximation decreases monotonically with $\gamma_1$, while it remains almost independent of the other parameters.
    Vertical dashed lines mark the boundaries between the E, D, and S regimes. 
    (c) Same data as in (b) but at a longer evolution time ($\omega t=10$), highlighting how the D phase is the hardest to capture correctly.}\label{fig:error}
\end{figure*}

To monitor the quality of the approximation of the target dynamics, we compute a time-averaged approximation ratio on the local magnetization $\sigma^z_j$:
\begin{equation}
 \mathcal{E}(t) = 1-\frac{\int_0^t {\rm d}t' \sum_j \left( \braket{\sigma^z_j(t')}_{\rm ex}-\braket{\sigma^z_j(t')}_{\rm PS}\right)^2}{\int_0^t {\rm d}t' \sum_j \left( \braket{\sigma^z_j(t')}_{\rm ex}\right)^2} \ .
\end{equation}
The labels $\braket{\sigma^z_j(t')}_{\rm ex}$ and $\braket{\sigma^z_j(t')}_{\rm PS}$ mark the expectation value of the magnetization at time $t'$ with the exact non-Hermitian dynamics of the two-level systems and the PS dynamics on the full qutrit model, respectively.
A value of $ \mathcal{E}(t) = 1$ means the evolution of the local observable is perfectly reproduced, while $ \mathcal{E}(t) = 0$ corresponds to a trivial (flat) dynamics of $\braket{\sigma^z_j(t)}_{\rm PS}$.
In the worst case, $\mathcal{E}(t)$ can become negative, indicating that the approximation is even worse than a trivial infinite-temperature average.
We note that $\mathcal{E}(t)$ can be a rather severe figure of merit, e.g., if one is interested in qualitative features such as oscillation periods rather than precise amplitudes.

In our data, we do not see a clear universal behaviour emerging in $\mathcal{E}(t)$; rather, the decrease in the approximation accuracy depends nontrivially on the targeted point in the phase diagram and on the system size. 
We summarize some of its features in Fig.~\ref{fig:error}.
If we fix the system size and the two-body dissipation rate $\gamma_2$, we observe a general decrease of the accuracy in time, as reported in panel (a) for $N=8$ and $\gamma_2=0.8\omega$, after a plateau around $\mathcal{E}(t)\sim 1$ for relatively short evolutions. 
The time frame at which the PS dynamics starts deviating from the target one depends on $\gamma_1$, with clear qualitative differences between the three phases.
In E, the agreement is excellent at short to intermediate times, while at long times $\mathcal{E}(t)$ decreases almost linearly because of the progressive mismatch between the oscillation amplitude of the PS and exact dynamics.
In D, instead, we see a sharp drop in accuracy after the transient (see inset), due to the dampening of the oscillations in the PS dynamics, in contrast to the persistent ones in the target evolution; see Fig.~\ref{fig:N2_phenomenology}(c) for a qualitative understanding. After that, $\mathcal{E}(t)$ saturates because the PS qutrit evolution does capture the constant offset around which $\braket{\sigma^z_j(t)}$ oscillates.
Finally, the S phase displays some deviations at short times, due to minor differences in the transient, but then the approximation is almost perfect as both exact and approximated evolution reach the trivial steady state at long times.

For a better understanding of the role played by the dissipation rates and the system size at different time scales, we plot in Fig.~\ref{fig:error}(b-c) two snapshots of the approximation ratio $\mathcal{E}(t)$ at two different times as a function of $\gamma_1$, for different choices of $\gamma_2$ and $N$.
At short times $\omega t=2$, corresponding to the plateau in panel (a), both $N$ and $\gamma_2$ play a minimal role. Instead,  $\mathcal{E}(t)$ decreases monotonically with $\gamma_1=\Omega^2 \Delta t/2$: this is expected, as a stronger coupling $\Omega$ to the auxiliary space makes larger-order terms that are not described by Eq.~\eqref{eq:nh_approx} more relevant. 
However, this effect remains minimal, since the deviations from unity are less than 1\%.
At larger times, see Fig.~\ref{fig:error}(c), the overall behaviour of $\mathcal{E}$ is less regular, but it clearly shows that the D phase is the hardest to capture and the most sensitive to the value of $\gamma_2$ and $N$. Indeed, increasing $\gamma_1$ in the E region leads to a worse approximation as the system approaches D, but then the accuracy increases again in the trivial phase S, as one can also appreciate from panel (a).
Counterintuitively, the simulation accuracy increases for larger two-body dissipation $\gamma_2$, where one may expect the second-order approximation in $\Delta t$ behind Eq.~\eqref{eq:nh_approx} to become less accurate due to the increased relevance of higher-order terms.

The second limitation of the PS protocol originates from the inherent stochasticity of the measurement outcomes that randomly project on either the computational subspace $P$ or its auxiliary complement $Q$ with a probability proportional to the weight of the wavefunction in each sector.
Hence, in the PS dynamics, we expect to retain a fraction of trajectories that is exponentially small in the system size $N$ and the simulation time $t$; we denote this quantity as $p_{\rm PS} \simeq \nep^{-\alpha N t}$. The exponential dependence in $N$ and $t$ originates from the finite probability of each qutrit, encoded by the value of $\alpha$, to be projected onto the auxiliary subspace at each measurement. The probability of {\em never} doing so, therefore, is exponentially suppressed in the quantum volume $Nt$ of the circuit.

\begin{figure}
    \centering
    \includegraphics[width=\linewidth]{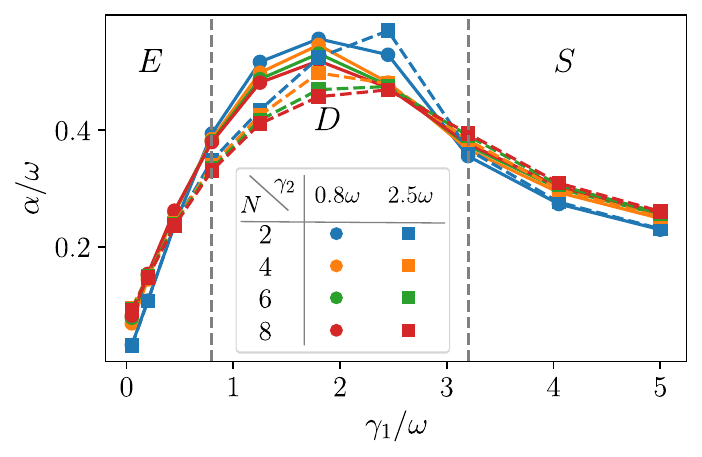}
    \caption{Decay rate per particle of the success probability of post-selection as a function of $\gamma_1$. Data for $\omega \Delta t=0.1$. Circles with solid lines: $J=4\omega\ (\gamma_2 =0.8\omega )$; squares with dashed lines: $J=7\omega\ (\gamma_2 =2.45\omega )$. Vertical dashed lines separate the three phases. The small residual size dependence confirms the overall exponential suppression in $N$.}
    \label{fig:decay_rate}
\end{figure}

To efficiently characterize this decay without simulating a large number of stochastic trajectories, we force the measurement to project the system onto the computational space $P$ and keep track of the norm loss after each such projection. The latter is directly related to the probability of ``leaking out'' from $P$ to the auxiliary subspace $Q$.
In practice, we approximate the probability of successful postselection at the time step $t_m$ as
\begin{equation}\label{eq:rate_effective}
    p_{\rm PS}(t_m) \simeq \prod_{m'<m} \braket{\psi_{t_m'}|P|\psi_{t_m'}} \ ,
\end{equation}
where $\ket{\psi_{t_m'}}$ is the normalized state after $m'$ steps of the PS evolution.
In Fig.~\ref{fig:decay_rate}, we report the rate $\alpha$ obtained from fitting Eq.~\eqref{eq:rate_effective} with an exponential decay.
The small $N$ dependency suggests that the exponential \emph{Ansatz} with decay rate $\alpha$ correctly captures the leading size scaling. 

The rate strongly depends on $\gamma_1$, as it drives the system through the different regions of the phase diagram, while $\gamma_2$ seems to play only a minor role, as for the considered parameters it does not incur any crossing of a phase boundary.  
The two phases D and S are severely hampered by the fast exponential decay; combining this with the approximation accuracy makes the degenerate phase the hardest to access on a digital quantum simulator with our protocol outside the initial transient regime.
In contrast, the small $\alpha$ in E, achieved somewhat counter-intuitively at a large two-body dissipation rate $\gamma_2$, allows for probing the entangled phase up to comparatively large system sizes and time scales.

As observed also in the simulation accuracy, see Fig.~\ref{fig:error}, in the E and D regions, increasing the value of $\gamma_2$ helps in reaching a better performance. Not only does a large two-body dissipation lead to higher accuracy, it also decreases the decay rate of the fraction of PS trajectory, even though only mildly. 
This suggests that the most interesting phenomenology of the model, i.e., the correlated steady state in E and the persistent oscillations between multiple steady states in D, can indeed be probed with our protocol at large effective non-Hermitian interaction, possibly with an exponential but still manageable overhead in the trajectory statistics. 

\section{Conclusions}\label{sec:conclusion}
In this paper, we provided an explicit protocol to engineer tunable non-Hermitian models in a qutrit-based digital quantum simulator, harnessing the quantum Zeno effect. By alternating unitary evolution with frequent stroboscopic projective measurements on a single level of the qutrits, considered as an ancillary subspace, we induce an effective non-Hermitian Hamiltonian onto the spin-$\frac{1}{2}$ Hilbert space spanned by the remaining levels. 
This minimal setup allows for a high flexibility in the design of the non-Hermitian terms, which depend on the connectivity and the unitary operations available on the qutrit setup and can be both one-body or $k$-body terms.

As a benchmark model, we focused on a non-Hermitian spin-$\frac{1}{2}$ system with both local dissipation and nearest-neighbour density--density interaction terms, originating from a combination of Rabi oscillations and Ising coupling involving the monitored level. Despite the apparent simplicity, this model displays nontrivial correlated phases already at small sizes~\cite{zeni2025theorycorrelatedquantumzeno}, making it an ideal test of the proposed PS digital quantum simulation. Indeed, we can capture the physics of all three different phases of the model, even though a quantitative agreement depends more subtly on the model parameters and system size.
Other complex interacting non-Hermitian Hamiltonians can be designed as well, as we briefly discuss in Appendix~\ref{app:extra_models}. 

Here, we only focused on effective spin-$\frac{1}{2}$ non-Hermitian Hamiltonians, but our approach readily generalizes to multi-level systems. The minimal setting is to use $d$-dimensional quantum information carriers to design effective Hamiltonians with $d-1$-dimensional local Hilbert spaces, as we did in this work. However, more intricate measurement protocols involving multiple auxiliary levels could further extend the class of non-Hermitian terms that can be designed, opening exciting prospects for future investigations.

The main limitation concerns the strength of the non-Hermitian terms in the effective Hamiltonian: the stronger they are, the farther the time evolution is from the quantum Zeno regimes where the approximations behind the effective dynamics are well under control. This generically leads to a worse simulation accuracy of the target non-Hermitian model and a smaller post-selection success rate.
On top of these, there will be the hardware and measurement noise to take into account. The impact of the first strongly depends on the target dynamics and how fragile the transient and the steady state are to further decoherence. The measurement noise, instead, is inherently more problematic as it will create a number of ``false positives'' hindering the post selection and requiring an even larger number of runs.
However, the benchmark model shows that the main qualitative features are accessible at a strong two-body dissipation rate, at least in the transient regime, making it experimentally feasible at least on small qutrit chains.

An important aspect that we did not address in this work is the role of the measurement distributions. Here, we assumed that all qutrits are measured at regular intervals, but one can introduce further stochasticity by fixing the monitoring rate and choosing randomly which qutrit to measure at each step. The two approaches have the same ensemble average in the continuous-time limit, but the trajectory distribution is different, possibly reflecting a different approximation of the underlying effective non-Hermitian Hamiltonian~\cite{Wauters_PRB2025}.

\section*{Data availability statement}
All data supporting this paper can be found in Ref.~\cite{wauters_2026_20828960}.

\begin{acknowledgements}
This project has received funding from the European Union’s Horizon Europe research and innovation programme under grant agreement No 101080086 NeQST, from the Swiss State Secretariat for Education, Research and Innovation (SERI) under contract number UeMO19-5.1.
A.B. acknowledges financial support by the European Union—NextGeneration EU, within PRIN 2022, PNRR M4C2, TANQU Project No. 2022FLSPAJ (CUP Grant No. B53D23005130006).
P.H. has received additional funding from the European Union under NextGenerationEU, PRIN 2022 Prot. n. 2022ATM8FY (CUP: E53D23002240006).
Views and opinions expressed are however those of the author(s) only and do not necessarily reflect those of the European Union or the European Commission. Neither the European Union nor the granting authority can be held responsible for them.
This work was supported by the Provincia Autonoma di Trento and Q@TN, the joint lab between the University of Trento, FBK—Fondazione Bruno Kessler, INFN—National Institute for Nuclear Physics, and CNR—National Research Council.

\end{acknowledgements}

\appendix
\section{Odd $N$}\label{app:oddN}
\begin{figure}
    \centering
    \includegraphics[width=\linewidth]{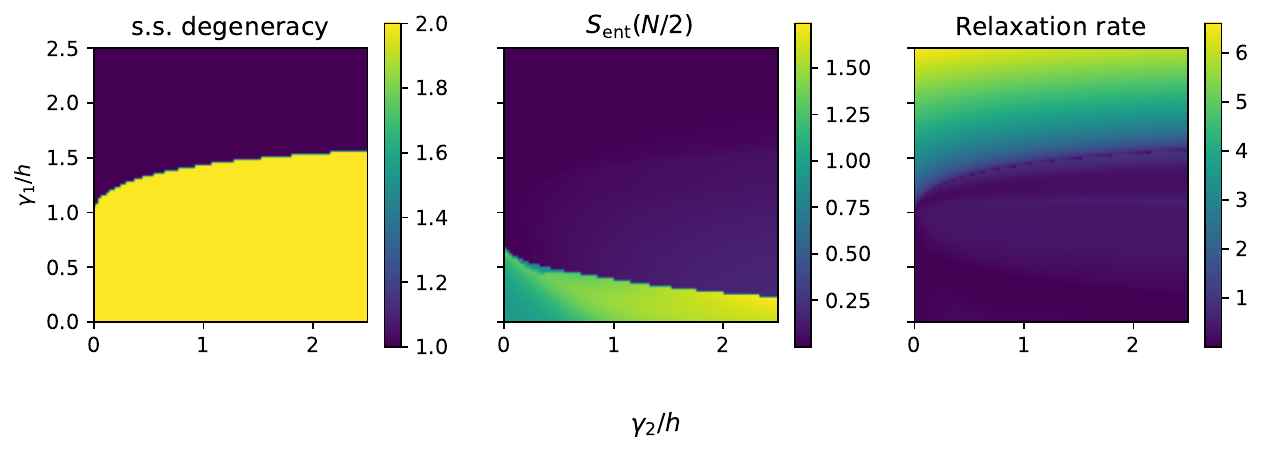}
    \caption{Ground state degeneracy (left), entanglement entropy (center), and relaxation rate (right) for the effective non-Hermitian qubit model with $N=7$. The values of $\gamma_1$ and $\gamma_2$ define $3$ different phases of the system: separable degenerate phase S-D (double degeneracy of the steady state, low entropy, and low relaxation rate), entangled degenerate phase E-D (double degeneracy, high entropy, and low relaxation rate), and standard phase S (single steady state, low entropy, and high relaxation rate).}
    \label{fig:phase_diagram_N7}
\end{figure}

The model investigated in the main text displays an interesting even--odd effect in the steady-state degeneracy, observable already at $N=3$.
Figure~\ref{fig:phase_diagram_N7} reports the figures of merit we chose to characterize the steady-state properties across the phase diagram of the model for $N=7$: the steady-state degeneracy (a), its entanglement (b), and the relaxation rate (c).
In stark contrast with the analogous data for $N=8$, see Fig.~\ref{fig:phase_diagram_N8}, the steady degenerate region extends all the way to $\gamma_1=0$. However, the entanglement structure of the degenerate steady states does depend on $\gamma_1$ still, and now we can distinguish between an entangled degenerate phase (E-D) and a separable degenerate one (S-D). 
It will be an interesting question for the future to study how far these even--odd effects persist in the thermodynamic limit.

\section{Further non-Hermitian models}\label{app:extra_models}
In this appendix, we provide further examples of non-Hermitian terms that one can engineer from the PS qutrit dynamics. These serve to illustrate the richness of models accessible by this approach. 

\subsection{Non-Hermitian East model}
The protocol we described in the main text is not restricted to purely dissipative non-Hermitian terms. By carefully combining different coupling terms between the qutrits, one can design more exotic Hamiltonians, such as a non-Hermitian version of the East model~\cite{Pancotti_PRX2020,das_2026}, describing (non-Hermitian) Rabi oscillations activated by the occupation on a neighbouring site
\begin{equation}\label{eq:East_nh}
    H_{\rm eff} = -i \lambda \sum_j n_j \sigma^x_{j+1}\,.
\end{equation}
To engineer it, we need three terms involving the auxiliary levels in the qutrit Hamiltonian
\begin{equation}
    H_{\rm full} = \Omega_1 \sum_j X^{(1,2)}_j + \Omega_2\sum_j X^{(1,2)}_jX^{(0,1)}_{j+1} + \Omega_3 \sum_j X^{(0,2)}_j \ .
\end{equation}
After projecting the evolution operator onto the $P$ subspace and setting $\Omega_3^2=\Omega_1^2+\Omega_2^2$, we get Eq.~\eqref{eq:East_nh}  with $\lambda = \Omega_1 \Omega_2 \Delta t$.

\subsection{Non-Hermitian Skin effect}
As a third example, we consider an effective Hamiltonian that displays the non-Hermitian skin effect~\cite{Zhang_NatComm2021,Zhang2022}. The simplest setting in which this effect appears is given by free-particle models with a non-reciprocal hopping term, leading to an exponential localization of all eigenstates on the edges of a one-dimensional chain with open boundaries. This extreme sensitivity to the boundary conditions is one of the hallmarks of non-Hermitian topology.

Let us consider the following Hamiltonian of the full qutrit system
\begin{eqnarray}\label{model_3_original}                                              
    H_{\rm full} &= J\sum_j \left(X_j^{(02)}Y_{j+1}^{(12)}+X_j^{(12)}X_{j+1}^{(02)} \right) \nonumber \\
    & +K\sum_j\left(X_j^{(01)}X_{j+1}^{(01)}+Y_j^{(01)}Y_{j+1}^{(01)} \right) \ ,
\end{eqnarray}
where the coupling within the system subspace (second line) is chosen to be dependent on the Trotter step $\Delta t$, i.e., we set $K=\frac{J^2\Delta t}{2}$. At each time step, we perform projective measurements of the auxiliary level as described in Sec.~\ref{sec:framework}.
Applying Eq.~\eqref{eq:nh_approx} then leads to
\begin{equation} \label{model_3_target}   
    H_\mathrm{eff}=J^2\Delta t \sum_j \sigma_j^{-}\sigma_{j+1}^{+}-i J^2 \Delta  t \sum_j\big(n_j+n_{j+1}-n_j n_{j+1} \big) \ .
\end{equation}
The first term is indeed a nonreciprocal hopping responsible for the skin effect. The remaining terms diagonal in the density consist of projectors that necessarily appear in our measurement-based approach.
If needed, local projectors on one state can be balanced out by adding further coupling terms that result in the complementary projector, such as $-iJ^2 \Delta t (1-n_j)$. Hence, the effective Hamiltonian would contain only a trivial population loss. Importantly, this does not alter the number of measurements nor the post-selecting procedure, while it can affect the success probability decay rate.

In this last example, the nonreciprocal hopping emerges from balancing the original unitary dynamics on the qutrits and the effective Hamiltonian obtained with the post-selection procedure. This trick increases the classes of non-Hermitian terms that we can simulate. Importantly, even though the number of unitary gates acting on the qutrit increases, the mid-circuit readout schedule is always the same, suggesting that there is no measurement overhead with respect to the simpler examples shown in the main text.

%

\end{document}